# Shot-Peening of Pre-Oxidized Plates of Zirconium: Influence of Residual Stress on Oxidation

Laura Raceanu[a], Virgil Optasanu[a], Tony Montesin[a], Guillaume Montay[b], Manuel François[b]

[a]*ICB UMR 6303 CNRS – Université de Bourgogne, 9 Av. Alain Savary, 21078 Dijon Cedex, France*
[b]*LASMIS - Université de Technologie de Troyes, 12 rue Marie Curie, 10010 Troyes Cedex, France*

[laura.raceanu@u-bourgogne.fr](laura.raceanu@u-bourgogne.fr); [virgil.optasanu@u-bourgogne.fr](virgil.optasanu@u-bourgogne.fr); [tony.montesin@u-bourgogne.fr](tony.montesin@u-bourgogne.fr); [guillaume.montay@utt.fr](guillaume.montay@utt.fr); [manuel.francois@utt.fr](manuel.francois@utt.fr)

**Abstract.** The present study deals with the oxidation behavior under residual stress of shot-peened plates of "commercially pure" Zirconium exposed for 30 min at 650°C. The influence of the shot-peening on a pre-oxidizedmaterial is presented. The results have been used to determine the influences of these chemical (pre-oxidation) and mechanical (shot-peening) treatments on the high temperature oxidation of Zirconium. The oxygen profile was revealed using micro-hardness techniques and confirmed by SEM-EDS techniques. After pre-oxidation the samples were first resurfaced then shot-peened in order to introduce residual stress. A significant increase of the hardness of about +400 HV was observed on pre-oxidized shot-peened samples. The thermogravimetric analysis for 30 min at 650°C under 200mbar $O_2$ shows a significantly slower oxidation rate for shot-peened samples.A comparison with our computations points out the role of the residual stresses on the diffusion and, consequently, on the oxidation.

## INTRODUCTION

The influence of mechanical stresses on corrosion is well known and many authors work on this topic. As early as 1935, Gorsky[1] started working on the influence of the stresses on chemical diffusion. Experimental findings have been widely documented. A first theoretical base for the coupling between stress and diffusion based on thermodynamics was proposed by Larché and Cahn [2,3] and includes the influence of the stress gradient on the interstitial diffusion. More recently [4-12], studies have developed the thermodynamic model and have completed the dependence of the diffusion with respect to the stress gradient and stress values.

The main objective of this paper is to show that diffusion can be controlled by the residual stress. In previous studies we showed that the nature of the stress and the shape of the spatial evolution of the stress values are of central importance. Could we influence and control the diffusion using "the right residual stress distribution" as a barrier or as a motor?

Several types of processes to introduce residual stress can be considered. In this study we opted for the shot-peeningbecause of the specific spatial distribution of the residual stress associated with this treatment. The metal considered here, as a substrate, is the Zirconium. The Zirconium oxidation is anionic. The diffusion of the oxygen in this metal is mainly grain-boundary diffusion under 600°C and interstitial volume diffusion below this temperature [13]. For this reason oxidation experiments have been carried out at 650°C.

# EXPERIMENTAL PROCEDURE

In order to study the influence of the residual stress on the diffusion and on the oxidation mechanisms at high temperatures, several tests have been realized. Shot-peening was used for introducing the residual stresses. This method was selected according to the results of the simulations of coupled stress-diffusion using the computational model presented in previous works [9] and briefly explained in the Computational Model section.

The experimental design is presented in **TABLE 1**. Three groups of samples were used in order to highlight our conclusions: shot-peened samples (Group I), pre-oxidized then shot-peened samples (Group II) and pre-oxidized, shot-peened and then oxidized samples (Group III). All samples were subjected to an initial 750°C / 2h / secondary vacuum recrystallization annealing. Rectangular 80 x 40 x 2 mm Zirconium plates have been used.

**TABLE 1.** Experimental design

| Groups | Treatment | No shot-peening | 10 min shot-peening | 30 min shot-peening |
|---|---|---|---|---|
| Group I | Shot-peening only | A | B | C |
| Group II | Pre-oxidation + shot-peening | D | E | F |
| Group III | Pre-oxidation + shot-peening + oxidation | G | H | I |
| Group IV | Shot-peening + Oxidation | J | K | L |

The sample A in Group I is the reference sample. Two different times of shot-peening are used: 10 and 30 min. The Group II is subjected to a particular preliminary pre-oxidation in order to proceed to a deep penetration of the oxygen in the substrate. A cycle alternating oxidation during 15 min at 200mbar $O_2$ atmosphere and 8h secondary vacuum maintain was used to produce a penetration of about 300µm of $O^{2-}$ atoms in the Zirconium substrate. This treatment also produced a layer of zirconia of 200µm thick. In order to proceed to a shot-peening on the oxygen enriched metal but not on the zirconia, the oxide layer was removed by resurfacing (precision grinding) as shown in **FIGURE 1**. In Group II residual stresses are introduced into a layer enriched with dissolved oxygen. Shot-peening parameters have been used to introduce residual stresses in a stratum of about 300µm thick under the free metal surface. The process uses tungsten carbide balls of 2 mm of diameter and a sonotrode vibrating at 20 kHz with an amplitude of 24µm.

One important fact for the interpretation of the results is that the grinding operation inevitably eliminates a layer of oxygen-enriched metal of about 100µm thick. The main consequence is that the maximum concentration of the dissolved oxygen in the metal at the free surface to be shot-peened is lower than the solubility of the oxygen in the metal.

Let's call the resurfaced side RS and the original non-resurfaced side NRS.

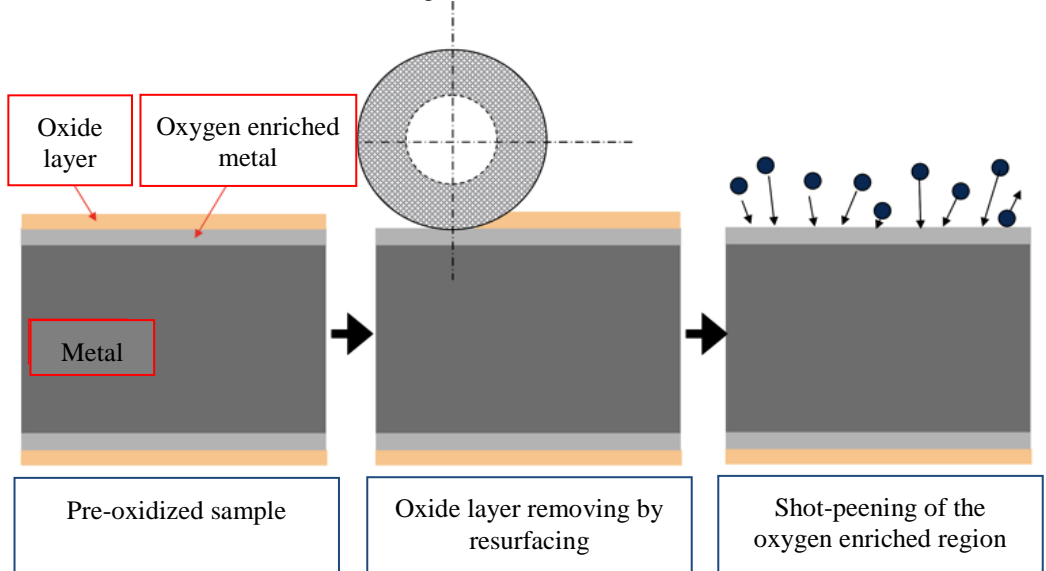

**FIGURE 1.** Surface preparation before shot-peening

The Group III initially undergoes the same chemical and mechanical treatment as Group II but is subjected to a supplementary oxidation of the samples in order to study the oxygen diffusion in a stressed, preliminary oxygen-enriched material. The oxidation was realized at 650°C, under 200 mbar $O_2$, during only 30 min in order to limit the mechanical stress relaxation by annealing. Oxidation duration and temperature were chosen as the best compromise between the necessitiesof havinga volume diffusion process and assuringthe persistence of mechanical stresses.

Parts of the samples B and C (shot-peened, without pre-oxidation) have been subjected to oxidation in a thermogravimetric balance in order to evaluate the influence of the shot-peening treatment on the mass gain. Let's call this group of samples Group IV.

The samples were analyzed by different techniques, namely:
- SEM with EDS and optical microscopy;
- Micro-hardness: each point on the hardness profiles has been obtained as an average of 6 measurements. The micro-hardness techniques have been used to reveal the presence of the oxygen in the substrate and/or the mechanical hardening given by the shot-peening;
- The incremental hole-drilling method have been used to measure the residual stress profile under the surface subject to the treatments. XRD technique was used to measure the value of the stresses at the samples surface and to validate the hole-drilling results;
- Thermogravimetric analysis (TGA).

## EXPERIMENTAL RESULTS

The influence of the shot-peening on the microstructure was revealed by optic microscopy and SEM.The samples were prepared for SEM by mirror polishing and chemical attacking with fluoric-nitric acid. The SEM images presented in**FIGURE 2**show the microstructure at different depthof the samples: under the shot-peened surface (a), at the core of the material (b) and near the opposite, non shot-peened surface (c). The grain structure was not affected by the mechanical treatment; there is no observable grain refinementor twinning. Therefore, the shot-peening introduces dislocations [14] inside the grains producing the hardening of the material and,consequently, residual stresses of second order, as a result of the high dislocation density.

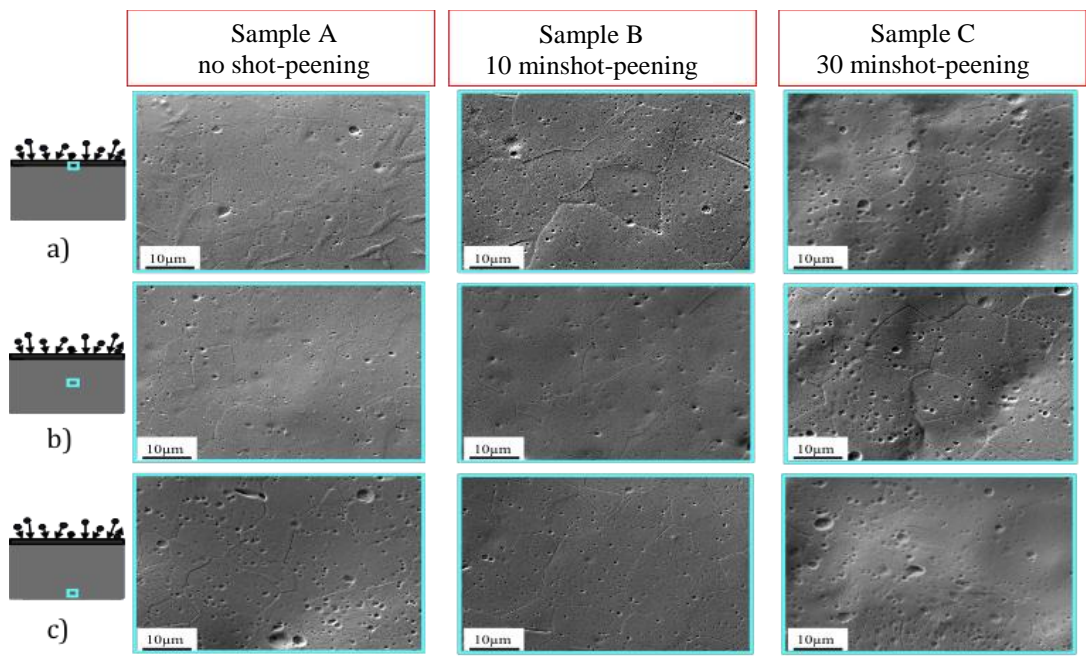

**FIGURE 2.**SEM images of shot-peened samples under the shot-peened surface (a), at the core of the material (b) and under the non-shot-peened surface (c).

The microscopy reveals the growth of the grain size from ~10µm to ~50 µm during the pre-oxidation process. This growth is due to the long treatment at high temperature. After pre-oxidation followed by grinding and shot-peening there is no observable modifications in the microstructure; there is no difference in the grain size or in grain morphology.

The micro-hardness profile on transversal sections was measured for all samples presented in **TABLE 1**. The samples were cold mounted and cut with a precision rotary disk cut-off machine. The transversal section was mirror-polished. The first point of measurement of the hardness was located at 20µm from the sample surface. Each point in the profile is the average value of 6 distinct measurements.

The comparison between the hardness profile and the composition obtained by EDS analysis shows a significant correlation (see **FIGURE 3**). The values of the oxygen concentration at the core of the material are not zero because of the oxygen enrichment after polishing. The measurements show the same thickness of the zone modified by the presence of the dissolved oxygen. As proved by Hayes et al. [14], micro-hardness measurements are good indicators of the size of oxygen-enriched zone. In our case the hardness is influenced by two factors: the dissolved oxygen and the high density of dislocations introduced by the mechanical treatment.

The shot-peening of an oxygen-free sample produces hardness increase, near the surface, of about 60 HV for the short treatment (10 min) and 75 HV for the long one (30 min). The hardness continuously decreases and reaches the value of the untreated material at about 300 µm from the surface. As remarked, there is no grain refinement, which implies that the increase of the hardness has two main origins: the increase of the density of dislocations and the residual stress.

The study of the hardness profile for pre-oxidized samples shows first of all that, despite the growth of the grain size from 10µm to 50µm, that the hardness value at the core of the material is not dependent on the grain size. The layers near the pre-oxidized surface are highly influenced by the presence of the dissolved oxygen. The presence of the oxygen produces a hardness increasesuperior to 400 HV compared to the reference sample. The resurfacing of the sample produces an ablation of a layer of 100 µm of oxygen-enriched metal, which means that the shot-peening is applied to a surface with ahardnessof about 300 HV. The shot-peening of the oxygen-enriched metal produces a hardness increase of about 130 HV as well a spread of the shape of the profile; the hardness is affected up to 300µm of depth.

A comparison of the hardness profilesof several samples is presented in**FIGURE 4**. The origin of the horizontal axis is taken at the position of the surface of the grinded samples.In order to take into account the ablation of the samples by the grinding operation, for the samples that were not resurfaced,the zero of the horizontal axis will be 100µm right shifted on the **FIGURE 4**.

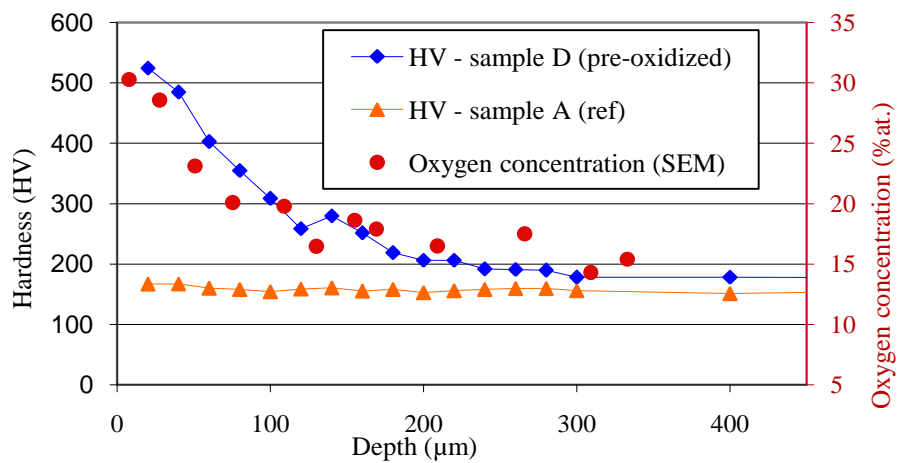

**FIGURE 3.** Correlation between hardness and oxygen-composition profiles

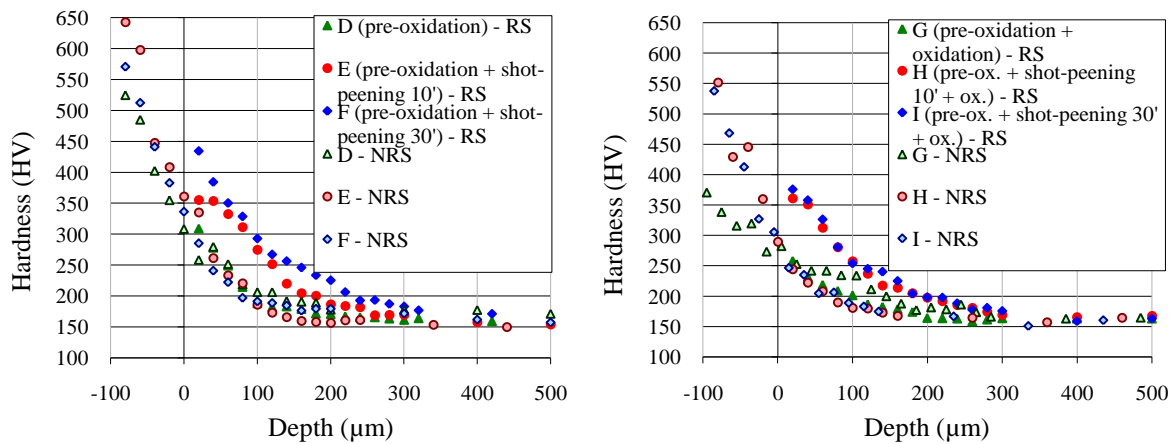

**FIGURE 4.**Comparison of hardness profile of Group II and Group III samples

A synthetic comparison between different hardness profiles for the resurfaced side of several samples is presented in **FIGURE 5**. The origin of the depth axis is the RS surface position. The main conclusions are:

- The insertion of oxygen produces a hardness increase more important than shot-peening (samples C and D);
- The shot-peening of an oxygen-enriched material is more efficient than the simple addition of the oxygen influence and the shot-peening influences (samples C, D and F);
- A particular shape of the hardness profile of shot-peened samples shows an inflexion of the curve between 100 μm and 200 μm. This inflexion is magnified on the sample "I" which was oxidized after the introduction of the residual stress by shot-peening. This particularity is not distinctly present on the others curves. We will attach a particular attention to this characteristic in the next paragraph;
- A comparison of samples F and I shows that the hardness is lower after the final oxidation process. This is due to the decrease of the dislocation density and the residual stress provoked by the annealing at this temperature.

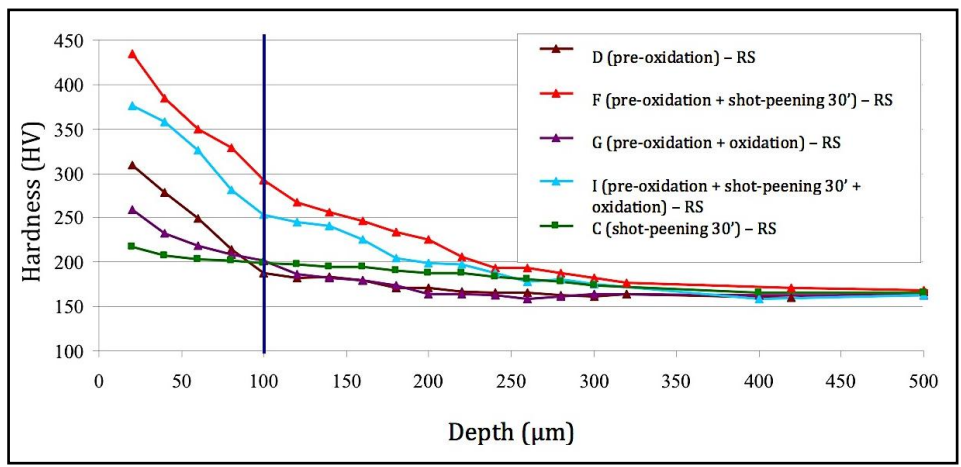

**FIGURE 5**. Hardness comparison of the resurfaced side (RS) for different sample treatments

The residual stresses have been measured by the incremental hole-drilling method. **FIGURE 6** presents an example of the residual stress profile for 10 min shot-peened samples with or without pre-oxidation. The measurements show a particular shape of the stress distribution: the most superficial layers are slightly compressed then deeper the negative stress gradient leads to a minimum algebraic value located at circa 100 μm of depth followed by an inversion of the gradient sign leading to positive values. This inversion of the stress gradient has an important impact on the oxygen flow. This aspect will be examined in the next section.

The stress gain given by the pre-oxidation is significant; the value of the maximal residual stress is almost doubled. The location of the peak of the stress profile is also shifted by about 30 μm towards the core of the material between samples B and E.

The residual stresses were not measured in the samples of the Group III samples because of the oxide layer produced after the final oxidation. For the F samples, the shape of the spatial distribution of the residual stress is very similar to those obtained for the E samples but the maximal compression value is higher, about 1GPa.

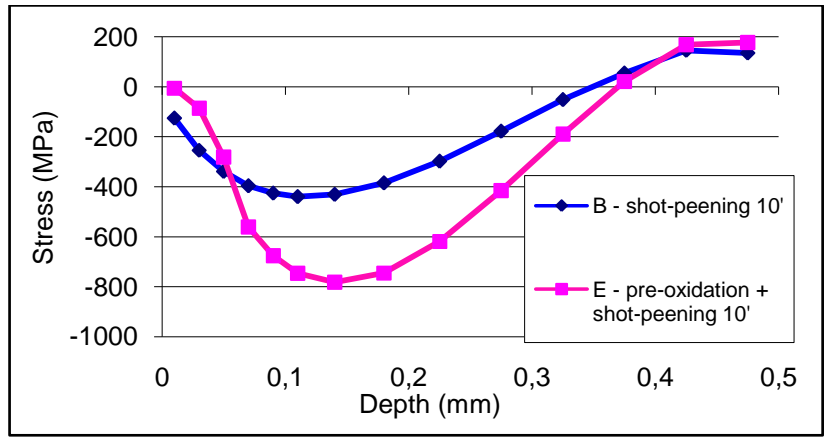

**FIGURE 6**. Residual stress introduced by shot-peening on preoxidated samples measured
by the incremental drilling method on samples B and E (10 min shot peening)

The values of the residual stress on the shot-peened surface were also measured by XRD. The results are concordant with those obtained using the incremental hole-drilling method.

The residual stresses introduced by the shot-peening were used subsequently in computations, to estimate the theoretic diffusion into the samples surfaces affected by the process.

In order to evaluate the influence of the residual stress on the oxidation rate, the mass gain of shot-peened samples was measured using the TGA technique. Parts of the samples B and C (10 x 10 x 2 mm) have been oxidized during 30 min under 200 mbar $O_2$. The results show very different behavior between the reference sample and the shot-peened samples. Despite the fact that the samples were shot-peened on one side only (36% of the total surface if the lateral faces are considered as unaffected) the differences in the mass gain are 12% for sample B and 40% for sample C. This result is significant. We suppose that the residual stresses are partially responsible for this behavior. The superficial layer of the shot-peened face presents no grain refinement, but a high density of dislocations. The dislocations are, normally, short circuits of diffusion, they should increase the oxidation rate. Slowing the rate of oxidation can thus be explained by the mechanical stresses. This aspect is developed in the computational model section.

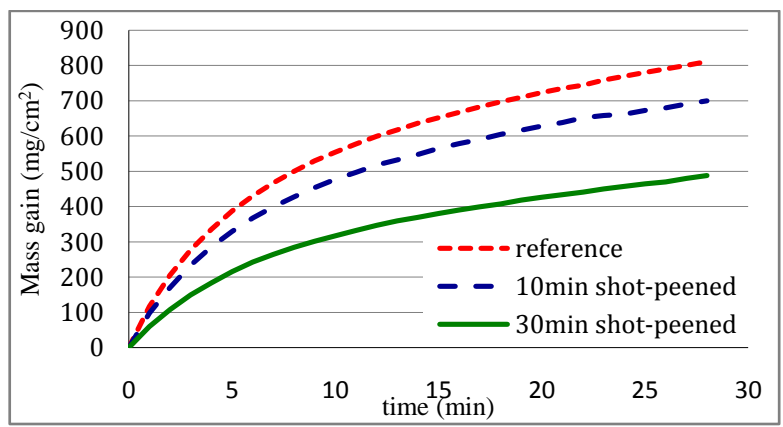

**FIGURE 7.** Mass gain of shot-peened samples (Group IV) during oxidation at 650°C

Han et al. [15] studied the corrosion behavior of the zirconium after a SMAT (surface mechanical attrition treatment) process that produces a grain refinement under the mechanically treated surface. They conclude that the corrosion resistance of Zr tends to increase with the reduction of grain size, which is related to the dilution of segregated impurities at grain boundaries as a result of grain refinement and the formation of passive protection film. In our case there is no grain refinement.

The slowing of the oxidation rate on a shot-peened sample could have a complex set of causes. One of the most important is that the shot-peening could influence the oxidation rate by changing the oxygen adsorption conditions by polluting the extreme surface layers. We think that this pollution could only change the very beginning of the oxidation process. After 30 min of oxidation the oxide layer is about 4 μm thick, which is widely superior to the layer eventually affected by the pollution.

# COMPUTATIONAL MODEL

Over the last decade the authors have developed a coupled model between diffusionand mechanical stress,whichhas been previously presented elsewhere [4-12]. The thermodynamic background has already been presented and the very final form of the model was recently submitted for publication [16]. The mainfeature of the model is to describe the bulk anionic diffusion of species in a substrate by taking into account the influence of the mechanical stresses and concomitantly the feedback of the oxygen concentration field on the mechanical state. The mechanical stressesconsidered here have several origins: the mechanical loads, the residual stresses and the self-induced stressproduced by the oxygen dissolution in the material. For this study, the model of coupling diffusion can be resumed by the expression of the oxygen flux in the zirconium:

$$\vec{J} = -D_0 \vec{\nabla} c - D_0 \frac{M_0}{RT} h_{ii} S_{ii} \vec{\nabla} c + D_0 \frac{M_0}{RT} c \vec{\nabla} (h_{ii} S_{ii}) \qquad (1)$$

with $c$ the oxygen concentration, $D_0$ the diffusion coefficient, $h_{ii} = \frac{\partial e_{ij}^{chem}}{\partial c}$ the chemical expansion coefficient, $M_0$ the molar mass of the diffusing species, $R$ the universal gas constant, $T$ the temperature and $S_{ij}$ the stress tensor.

This formulation shows the influences on the oxygen concentration profile. In our previous works we extensively developed the influence of the stress gradient and stress value on different materials and for different spatial distributions of the stress.It has been shown that the mechanical stress can slow down or accelerate the diffusion depending on the nature of the stress (compression or traction) as well the stress gradient.

In this simulation the residual stress has been taken into account and modeled in order to match the shape of the measured stress on the sample F (pre-oxidized and shot-peened 30'), as presented **FIGURE 8**.The computation does not simulate the oxidation but only the diffusion of $O^{2-}$ into the bulk of the metal; the frontier between oxide and metal is considered immobile. The spatial distribution of the residual stress has the following specificity:

- The beginning of the curve presented in**FIGURE 6**presents a negative stress gradient regionfallowed by a positive stress gradient zone. In the first zone, as can be seenin the Equation (1), the negative stress gradient term will give an upwind flow contribution in the matter flux, while in the second zone, the positive stress gradient will give an acceleration of the diffusion.
- The residual stress value is negative in the zone where the oxygen is present. The second term of the Equation (1) is then negative (because the concentration gradient is negative). This term will also give a negative contribution to the oxygen flux.

The computations simulate a long period of diffusion ($10^7$s), which is significantly higher than the actual time of the final oxidation process (1800s). This deliberate choice was made in order adequately to highlight the influences of stress on the diffusion and to draw clear conclusions.The influence of different terms in Equation (1) can be studied separately. **FIGURE 8** presents respectively: the initial concentration (corresponding to the concentration profileexperimentally obtained after the pre-oxidation process); the concentration profile computed with simple Fick's law;those computed using the stress gradient term; and those using the complete formulation of the coupling model. For shorter periods this effects are less evident but still present.

The formulation using only the stress gradient shows a reverse flow in the region where the stress gradient is negative. This strong effect affects the rest of the profile and, viewed globally, the concentration profile is below the curve given by a diffusion governed by Fick's law. The complete formulation presented in Equation (1) shows an inflexion of the concentration profile similar to those observed on the hardness profile and indicates that the influence of the stress is significant. This is an important result and tends to show that the oxidation rate can beinfluenced by the residual stress. The slowing of the oxidation rate can be influenced by several parameters,such as the oxygen adsorption and the pollution of the surface during shot-peening. However, the particular form of the oxygen distribution in the material would appear to be produced by the stress state.

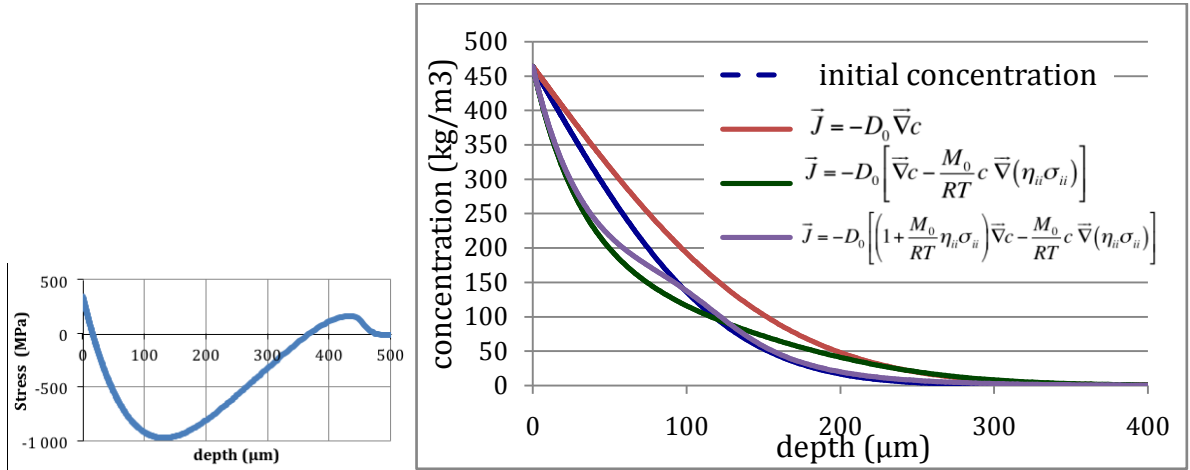

**FIGURE 8**. Stress profile and results of the computation (after $10^7$s of diffusion) for different terms of the Equation (1)

## CONCLUSION

The experimental study presented here deals with the oxidation of Zirconium samples, which are subjected to shot-peeningafter a pre-oxidation process.

The grain size is not affected by the hardening introduced by the shot-peening of the zirconium samples.

The comparison between experiments and computations shows similarities between the hardness profile and the simulated concentration profile. The particular form of the profiles, an inflexion point, is located in similar positions. This similarity strongly suggests that the residual stresses are the main factor of influence on the shape of the concentration profile. We have to remember here that the diffusion time used for our computations is several orders of magnitude larger than the real oxidation time. The experiments could not be processed over such a longtime for several reasons: the oxidation is metal consuming and the long exposure of the pre-stressed Zirconium at high temperatures eliminates of the residual stress. However, the computations show a trend, which seems to be magnified in the experimental process.

The thermogravimetric analysis of shot-peened samples shows that the mass gain decreases with the shot-peening time. This decrease is thought to be provoked by the residual stress, as is highlightedby the computations. The negative residual stress gradient, into the thin layers close to the surface acts as a reverse driving force. The nature of the residual stress (compressive stress) in the peened region acts too as a reverse driving force. Thus, the slowing of the oxidation rate can be explained, on the first hand by the negative residual stress gradient on the beginning of the stress distribution curve and, on the other hand, by the fact that the residual stress is compressive, which implies that the flux equation is influenced by two terms in opposition with the classic Fick's term.

The main conclusion of this study is that the diffusion can be influenced by the right choice of the process that introduces the residual stress. More the gradient of the residual stress and/or the value of the compressive stress are important, more the slowing effect of the stress is significant.The experimental results will be completed by other studies concerning the adsorption after the shot-peening of the metal surface and about the pollution introduced by the tungsten carbide balls used in the process. Other experiments would need to be devised to deal with lower temperatures - in order to maintain the stress over a long period of time. The nitriding process is a good candidate. In addition, the diffusion of hydrogen in metals could provide a good opportunity to influence or control the diffusion using the residual stress.